\documentclass[prd,showpacs,preprintnumbers,amsmath,amssymb]{revtex4}

\usepackage{graphicx}
\usepackage{dcolumn}
\usepackage{bm}

\begin{document}

\preprint{hep-ph/0011227}

\title{Emergence of Classicality in Quantum Phase Transitions}

\author{Sang Pyo Kim}\email{spkim@phys.ualberta.ca;
sangkim@ks.kunsan.ac.kr} \affiliation{Theoretical Physics
Institute, Department of Physics,
University of Alberta, Edmonton, Alberta, Canada T6G 2J1\\
and Department of Physics, Kunsan National University, Kunsan
573-701, Korea}

\author{Chul H. Lee} \email{chlee@phys.ualberta.ca; chlee@hepth.hanyang.ac.kr}

\affiliation{Theoretical Physics Institute, Department of Physics,
University of Alberta, Edmonton, Alberta, Canada T6G 2J1\\ and
Department of Physics, Hanyang University, Seoul 133-791, Korea}

\date{\today}

\begin{abstract}
We show that the long wavelength modes of a field become classical
during a second order phase transition because of the interaction
with the short wavelength modes of the field. In a massive scalar
frield model the number and thermal states of long wavelength
modes, whose Wigner functions are sharply peaked around the
classical trajectories during the phase transition, exhibit only
classical correlation without achieving quantum decoherence. In a
linearly coupled scalar field model, the long wavelength modes are
shown to effectively achieve quantum decoherence because of the
mode-mixing. Finally we define a quantal ordering parameter that
is linear in the field variable and satisfies the classical field
equation.

\end{abstract}


\maketitle

\section{Introduction}

A system becomes classical when it recovers classical correlation
and loses quantum coherence, i.e. decoheres. The present Universe,
for instance, is believed to experience a quantum-to-classical
transition at a certain stage of its evolution from its early
quantum state. One of such decoherence mechanisms is the
interaction of the system with an environment
\cite{zurek,joos,caldeira,han}. The environment-induced
decoherence mechanism explains correctly the density perturbation
necessary for structure formation of the present Universe
\cite{kiefer}. A phase transition is another interesting
phenomenon in which one observes the quantum-to-classical
transition. For instance, the system of the long wavelength modes
of a self-interacting scalar field  become classical through the
interaction with the environment of the short wavelength modes
\cite{lombardo}.

Another aspects of classicality is classical correlation. The
parametric interaction of an open system  provides classical
correlation without any direct interaction with the environment.
The coupling constants (parameters) of the open system change
explicitly in time. A quantum field model in which the mass
parameter changes sign during a quenched second order phase
transition provides such an open system. In this model the
dynamical evolution of phase transition is mainly described by a
classical order parameter, whose quantum fluctuations are
necessary for domains or topological defects \cite{boyanovsky}. In
a previous paper \cite{kim-lee} (hereafter referred to I), we
introduced a quantum phase transition model without any classical
order parameter in which a certain symmetric quantum state drives
the phase transition. A question is then raised how the quantum
phase transition exhibits classical features.

In the field model a classical background field is introduced as
the order parameter, around which quantum field fluctuates
\cite{boyanovsky}. The expectation value of the field, however,
vanishes globally from the symmetry (parity) argument
\cite{lombardo}. There is still a room for explanation of the
order parameter in quantum theory: the condensation of the field
into a coherent state having a nonzero expectation value. However,
in Ref. I this initial quantum state is shown to be either a
coherent or a coherent-thermal state. If the system starts
initially from an exactly symmetric state such as the Gaussian
vacuum or thermal equilibrium, the quantum law yields zero
expectation value throughout the phase transition. On the other
hand, in Ref. I another possibility is discussed that the phase
transition may proceed quantum mechanically in the symmetric
quantum state, such as the Gaussian vacuum, number, and thermal
states, even without globally forming a coherent condensate.

In this paper we shall show how classicality emerges from the
symmetric quantum evolution of phase transition. Our model is a
massive scalar field, the mass of which changes sign to mimick a
quenched second order phase transition. This model initially has
been used to study the behavior of an inflaton during the
slow-rollover in the new inflationary scenario \cite{guth} and
also has been employed to describe quantum processes of the phase
transition during the spinodal instability regime in Ref. I. It is
known that the Gaussian state of the model obeys a classical
probability distribution \cite{guth}. In this paper we shall apply
the quantitative measure of both classical correlation and quantum
decoherence in Ref. \cite{anderson} to the quantum phase
transition. For that purpose we derive the density matrices and
Wigner functions for the Gaussian vacuum, number state, and
thermal equilibrium.

It is found that the density matrix and Wigner function manifestly
exhibit classical correlation for the long wavelength (soft) modes
that grow exponentially, whereas both long and short wavelength
(hard) modes maintain the same initial quantum coherence. Hence
the massive scalar field does not achieve classicality in the
genuine sense. To show how quantum decoherence occurs during the
phase transition, we consider an analytically solvable model,
motivated by a self-interacting scalar field, in which a long
wavelength mode is linearly coupled to a short wavelength mode. It
is shown that the mode-mixing (coupling) between the long and
short wavelength modes drastically reduces the degree of quantum
coherence of the long wavelength mode during the phase transition.
Thus the long wavelength mode becomes completely classical from
the view point of both classical correlation and quantum
decoherence. It is further suggested that a quantal quantity can
be defined out of long wavelength modes to play a similar role of
the order parameter.

The organization of this paper is as follows. In Sec. II, we
derive the density matrices and the Wigner functions of a massive
scalar field in the Gaussian vacuum, number state, and thermal
equilibrium. In Sec. III, using the quantitative measure of
classical correlation and quantum decoherence, we show that the
unstable long wavelength modes exhibit classical correlation,
whereas the short wavelength modes maintain their initial quantum
coherence. In Sec. IV, we show how a long wavelength mode can
achieve quantum decoherence during the phase transition through a
linear coupling with a short wavelength mode. In Sec. V, we define
a quantal order parameter that is linear in the field variable and
satisfies the classical field equation.

\section{Density Matrix and Wigner Function}

As the first model for the second order phase transition, we
consider a real massive scalar field with the Lagrangian density
\begin{equation}
{\cal L} =  \frac{1}{2} [\dot{\phi}^2 - (\nabla \phi)^2 ] -
\frac{1}{2} m^2 (t) \phi^2, \label{act}
\end{equation}
where the mass is assumed to change sign during the phase
transition. Following Refs. I and \cite{kim1}, the action
(\ref{act}), upon a suitable mode-decomposition, leads to the
Hamiltonian
\begin{equation}
H (t) = \sum_{\alpha} \Biggl[\frac{1}{2} \pi_{\alpha}^2 (t) +
\frac{1}{2} \omega^2_{\alpha} (t)\phi_{\alpha}^2 (t) \Biggr] =
\sum_{\alpha} H_{\alpha} (t), \label{sc osc}
\end{equation}
where
\begin{equation}
\omega^2_{\alpha} (t) = m^2 (t) + {\bf k}^2, \label{freq}
\end{equation}
and $\alpha$ denotes the Fourier mode defined by
\begin{equation}
\phi_{\bf k}^{(+)} (t) = \frac{1}{2} [ \phi_{\bf k} (t) +
\phi_{-{\bf k}} (t)], \quad \phi_{\bf k}^{(-)} (t) =
\frac{i}{2} [ \phi_{\bf k} (t) - \phi_{-{\bf k}} (t)]. \label{fo mod}
\end{equation}
The Fourier modes ({\ref{fo mod}) are Hermitian because $\phi_{\bf
k}^* = \phi_{-{\bf k}}$. Thus the Hamiltonian (\ref{sc osc})
consists of the infinite number of decoupled, time-dependent
oscillators. In quantum field approach the fundamental law is the
functional Schr\"{o}dinger equation (in units of $\hbar = k = 1$)
\begin{equation}
i \frac{\partial}{\partial t} \Psi(\phi, t) = \hat{H} (t)
\Psi(\phi, t). \label{sch eq}
\end{equation}
As every mode is decoupled from each other, the wave functional to
Eq. (\ref{sch eq}) is now given by the product
\begin{equation}
\Psi(\phi, t) = \prod_{\alpha} \Psi_{\alpha} (\phi_{\alpha}, t)
\end{equation}
of the wave functions of each Schr\"{o}dinger equation
\begin{equation}
i \frac{\partial}{\partial t} \Psi_{\alpha}
(\phi_{\alpha}, t) = \hat{H}_{\alpha}
(t) \Psi_{\alpha} (\phi_{\alpha}, t). \label{mod sch eq}
\end{equation}
The Fock space consisting of exact quantum states for a
time-dependent oscillator was first constructed by Lewis and
Riesenfeld \cite{lewis} (for the references on many different
methods, see \cite{kim2}).

Now the quantum evolution of the Hamiltonian (\ref{sc osc}) is
reduced to that of individual oscillators in Eq. (\ref{mod sch
eq}). To find each Fock space, we follow the so-called
Liouville-von Neumann (LvN) approach developed in Refs. I and
\cite{kim1,kim2}. In the LvN approach one finds a pair of
operators for each mode of the Hamiltonian (\ref{sc osc}):
\begin{eqnarray}
\hat{a}_{\alpha} (t) &=&  i \Bigl[ \varphi_{\alpha}^* (t)
\hat{\pi}_{\alpha} - \dot{\varphi}_{\alpha}^*
(t)\hat{\phi}_{\alpha} \Bigr],
\nonumber\\\hat{a}_{\alpha}^{\dagger} (t) &=& - i \Bigl[
\varphi_{\alpha} (t) \hat{\pi}_{\alpha} - \dot{\varphi}_{\alpha}
(t)\hat{\phi}_{\alpha} \Bigr], \label{an-cr}
\end{eqnarray}
that satisfy the quantum LvN equation
\begin{eqnarray}
i \frac{\partial}{\partial t} \left\{ \begin{array}{c} \hat{a} (t)
\\ \hat{a}^{\dagger} (t) \end{array} \right\} + [ \left\{
\begin{array}{ll} \hat{a} (t) \\ \hat{a}^{\dagger} (t) \end{array}
\right\}, \hat{H}_{\alpha} (t)] = 0. \label{ln eq}
\end{eqnarray}
Equation (\ref{ln eq}) leads to the classical equation of motion
for a complex $\varphi_{\alpha}$:
\begin{equation}
\ddot{\varphi}_{\alpha} (t) + \omega^2_{\alpha} (t)
\varphi_{\alpha}(t) = 0. \label{cl sol}
\end{equation}
By requiring the Wronskian condition
\begin{equation}
\dot{\varphi}_{\alpha}^* (t) \varphi_{\alpha} (t) -
\dot{\varphi}_{\alpha} (t) \varphi_{\alpha}^* (t) = i,
\label{wron}
\end{equation}
one can make $\hat{a}_{\alpha} (t)$ and
$\hat{a}_{\alpha}^{\dagger} (t)$ satisfy the standard commutation
relation at any time
\begin{equation}
[\hat{a}_{\alpha} (t), \hat{a}_{\beta}^{\dagger} (t)] =
\delta_{\alpha \beta}.
\end{equation}
In the LvN approach the exact quantum state of the time-dependent
oscillator is determined, up to some time-dependent factor, by the
eigenstate of a Hermitian operator satisfying Eq. (\ref{ln eq}).
For instance, the vacuum state is the zero-particle state
$\hat{a}_{\alpha} (t) \vert 0, t \rangle = 0$ according to the
standard quantum mechanics. Hence, as the phase transition
proceeds, the time-dependent vacuum state at a late time is a
squeezed state of the initial vacuum state \cite{kim2,grishchuk}.

As both operators in Eq. (\ref{an-cr}) already satisfy Eq.
(\ref{ln eq}), we choose each number operator as
\begin{equation}
\hat{N}_{\alpha} (t) = \hat{a}_{\alpha}^{\dagger} (t)
\hat{a}_{\alpha} (t),
\end{equation}
and construct the Fock space consisting of the time-dependent
number state
\begin{equation}
\hat{N}_{\alpha} (t) \vert n_{\alpha}, t \rangle =
n_{\alpha} \vert n_{\alpha}, t \rangle.
\end{equation}
Then the vacuum state is given by
\begin{equation}
\hat{a}_{\alpha} (t) \vert 0_{\alpha}, t \rangle = 0,
\end{equation}
and the number state by
\begin{equation}
\vert n_{\alpha}, t \rangle = \frac{1}{\sqrt{n_{\alpha}!}}
[\hat{a}_{\alpha}^{\dagger} (t)]^{n_{\alpha}} \vert 0_{\alpha}, t
\rangle. \label{num st}
\end{equation}
In Ref. I the number state has the coordinate representation
\begin{equation}
\Psi_{n_{\alpha}} (\phi_{\alpha}, t) = \Biggl(\frac{1}{2 \pi
\varphi_{\alpha}^* \varphi_{\alpha}} \Biggr)^{1/4}
\frac{1}{\sqrt{2^{n_{\alpha}} n_{\alpha}!}}
\Biggl(\frac{\varphi_{\alpha}}{\varphi_{\alpha}^*}
\Biggr)^{n_{\alpha}/2} H_{n_{\alpha}}
\Biggl(\frac{\phi_{\alpha}}{\sqrt{2 \varphi_{\alpha}^*
\varphi_{\alpha}}} \Biggr) \exp \Biggl[\frac{i}{2}
\frac{\dot{\varphi}_{\alpha}^*}{\varphi_{\alpha}^*}
\phi_{\alpha}^2 \Biggr], \label{osc wav}
\end{equation}
where $H_{n_{\alpha}}$ is a Hermite polynomial. From the
definition $ \rho_{\Psi} (x', x) = \langle x' \vert \Psi \rangle
\langle \Psi \vert x \rangle$, we find the density matrix for the
number state (\ref{osc wav}):
\begin{eqnarray}
\rho_{n_{\alpha}} (\phi'_{\alpha}, \phi_{\alpha}, t) &=&
\Biggl(\frac{1}{2 \pi \varphi_{\alpha}^*
\varphi_{\alpha}}\Biggr)^{1/2} \frac{1}{2^{n_{\alpha}}
n_{\alpha}!} H_{n_{\alpha}} \Biggl(\frac{\phi'_{\alpha}}{\sqrt{2
\varphi_{\alpha}^* \varphi_{\alpha}}} \Biggr)
H_{n_{\alpha}} \Biggl(\frac{\phi_{\alpha}}{\sqrt{2
\varphi_{\alpha}^* \varphi_{\alpha} }} \Biggr) \nonumber\\
&& \times \exp \Biggl[- \frac{1}{2 \varphi_{\alpha}^*
\varphi_{\alpha}} \Biggl\{  \phi_{\alpha,C}^2 + \phi_{\alpha,\Delta}^2 \Biggr\} + i \frac{d}{dt} \ln (\varphi_{\alpha}^*
\varphi_{\alpha}) \phi_{\alpha, C} \phi_{\alpha,\Delta}
\Biggr], \label{den mat}
\end{eqnarray}
where
\begin{equation}
\phi_{\alpha,C} = \frac{1}{2} (\phi'_{\alpha} +
\phi_{\alpha}), \quad \phi_{\alpha,\Delta} = \frac{1}{2}(\phi'_{\alpha}
- \phi_{\alpha}).
\end{equation}
Once given the density matrix (\ref{den mat}), one easily finds
the Wigner function \cite{hillery}
\begin{equation}
P_{\alpha} (\phi_{\alpha}, \pi_{\alpha}) = \frac{1}{\pi} \int_{-
\infty}^{+ \infty} dy \langle \phi_{\alpha} - y \vert
\hat{\rho}_{\alpha} (t) \vert \phi_{\alpha} + y \rangle e^{2 i
\pi_{\alpha} y}. \label{wigner0}
\end{equation}
Substituting Eq. (\ref{den mat}) into Eq. (\ref{wigner0}), we
obtain the Wigner function for $\hat{H}_{\alpha}$
\begin{equation}
P_{n_{\alpha}} (\phi_{\alpha}, \pi_{\alpha}) = \frac{1}{\pi}
(-1)^{n_{\alpha}} L_{n_{\alpha}}(8 \varphi_{\alpha}^*
\varphi_{\alpha} \tilde{H}_{\alpha})e^{- 4 \varphi_{\alpha}^*
\varphi_{\alpha} \tilde{H}_{\alpha}}, \label{wigner}
\end{equation}
where $L_{n_{\alpha}}$ is a Laguerre polynomial and
\begin{equation}
\tilde{H}_{\alpha} (\phi_{\alpha}, \pi_{\alpha}) = \frac{1}{2}
\Biggl[ \pi_{\alpha} - \Biggl\{\frac{d}{dt} \ln( \varphi_{\alpha}^*
\varphi_{\alpha})^{1/2} \Biggr\} \phi_{\alpha} \Biggr]^2 + \frac{1}{8
(\varphi_{\alpha}^* \varphi_{\alpha})^2}\phi_{\alpha}^2.
\label{eff ham}
\end{equation}

We now turn to the density matrix and Wigner function for the
thermal state. In Refs. I and \cite{kim1}, the following density
operator is used for each mode
\begin{equation}
\hat{\rho}_{{\rm T} \alpha} (t) = 2\sinh \left(\frac{\beta
\omega_{\alpha, i}}{2} \right) e^{- \beta \omega_{\alpha, i}
(\hat{N}_{\alpha} (t) + \frac{1}{2})}, \label{den op}
\end{equation}
because $\hat{N}_{\alpha} (t)$ already satisfies the quantum LvN
equation. Here the free parameters $\beta$ and $\omega_{\alpha,
i}$ are to be identified with the temperature and frequency,
respectively, of the initial thermal equilibrium. From the density
matrix of Ref. I
\begin{eqnarray}
\rho_{{\rm T} \alpha} (\phi'_{\alpha}, \phi_{\alpha}, t) &=&
\left[\frac{\tanh\bigl(\beta \omega_{\alpha, i} /2 \bigr)}{2 \pi
\varphi_{\alpha}^* \varphi_{\alpha}}\right]^{1/2} \exp \Biggl[i
\frac{d}{dt} \ln (\varphi_{\alpha}^* \varphi_{\alpha})
\phi_{\alpha,C} \phi_{\alpha,\Delta} \Biggr] \nonumber\\ && \times
\exp \Biggl[-\frac{1}{2 \varphi_{\alpha}^* \varphi_{\alpha}}
\Biggl\{ \tanh\Biggl(\frac{\beta \omega_{\alpha,i} }{2} \Biggr)
\phi_{\alpha,C}^2 + \coth \Biggl(\frac{\beta \omega_{\alpha,i}
}{2} \Biggr) \phi_{\alpha,\Delta}^2 \Biggr\} \Biggr], \label{den
mat2}
\end{eqnarray}
we obtain the Wigner function
\begin{equation}
P_{{\rm T} \alpha} (\phi_{\alpha}, \pi_{\alpha}) = \frac{1}{\pi}
\tanh\Biggl(\frac{\beta \omega_{\alpha, i}}{2}\Biggr) \exp \Biggl[- 4
\tanh\Biggl(\frac{\beta \omega_{\alpha, i}}{2}\Biggr) \varphi_{\alpha}^*
\varphi_{\alpha} \tilde{H}_{\alpha} \Biggr]. \label{wigner2}
\end{equation}
Also, the density matrix (\ref{den mat}) and Wigner function
(\ref{wigner}) for the Gaussian vacuum state ($n_{\alpha} = 0$)
are obtained by taking the zero temperature limit $\beta
\rightarrow \infty$ of Eqs. (\ref{den mat2}) and (\ref{wigner2}).

The nature of the number state (\ref{osc wav}) and thermal state
(\ref{den op}) can also be understood from the Wigner functions
(\ref{wigner}) and (\ref{wigner2}). In particular, the 1-$\sigma$
contour of the Wigner function contains some useful kinematical
information of the quantum state in phase space. The 1-$\sigma$
contour of the thermal state is defined by
\begin{equation}
\Biggl\{\tanh\Biggl(\frac{\beta \omega_{\alpha,i}}{2}\Biggr)
\Biggr\} 4 \varphi_{\alpha}^* \varphi_{\alpha} \tilde{H}_{\alpha}
= 1. \label{1-sig}
\end{equation}
Here and hereafter the 1-$\sigma$ contour of the Gaussian vacuum
state is obtained by taking limit $\tanh (\beta \omega_{\alpha,
i}/2) =1$. As Eq. (\ref{1-sig}) describes a rotated ellipse in the
phase space $(\pi_{\alpha}, \phi_{\alpha})$, we can find the major
and minor axes through the transformation
\begin{eqnarray}
\tilde{\pi}_{\alpha} = \cos \theta_{\alpha} \pi_{\alpha} - \sin
\theta_{\alpha} \phi_{\alpha}, \nonumber\\ \tilde{\phi}_{\alpha} =
\sin \theta_{\alpha} \pi_{\alpha} +
 \cos \theta_{\alpha}
\phi_{\alpha},
\end{eqnarray}
where
\begin{eqnarray}
\tan 2 \theta_{\alpha} &=& \frac{2C_{\alpha}}{B^{+}_{\alpha} -
B^{-}_{\alpha}}, \nonumber\\ B^{+}_{\alpha} &=& \Biggl\{
\tanh\Biggl(\frac{\beta \omega_{\alpha,i}}{2}\Biggr) \Biggr\} 2
\varphi_{\alpha}^* \varphi_{\alpha}, \nonumber\\ C_{\alpha} &=&
\Biggl\{\tanh\Biggl(\frac{\beta \omega_{\alpha,i}}{2}\Biggr)
\Biggr\} \varphi_{\alpha}^* \varphi_{\alpha} \frac{d}{dt} \ln
(\varphi_{\alpha}^* \varphi_{\alpha}), \nonumber\\ B^{-}_{\alpha}
&=& \Biggl\{\tanh\Biggl(\frac{\beta
\omega_{\alpha,i}}{2}\Biggr)\Biggr\}
\Biggl[\frac{1}{2\varphi_{\alpha}^* \varphi_{\alpha}}+ \frac{1}{2}
\varphi_{\alpha}^* \varphi_{\alpha} \Biggl( \frac{d}{dt} \ln
(\varphi_{\alpha}^* \varphi_{\alpha}) \Biggr)^2 \Biggr].
\end{eqnarray}
Then Eq. (\ref{1-sig}) is written in a canonical form
\begin{equation}
\Biggl(\frac{\tilde{\pi}_{\alpha}}{\sqrt{1/\lambda^{+}_{\alpha}}}
\Biggr)^2 +
\Biggl(\frac{\tilde{\phi}_{\alpha}}{\sqrt{1/\lambda^{-}_{\alpha}}}
\Biggr)^2 = 1,
\end{equation}
where
\begin{equation}
\frac{1}{\lambda^{\pm}_{\alpha}} = \Biggl(\frac{B^{+}_{\alpha} +
B^{-}_{\alpha}}{2} \Biggr) \pm \Biggl( \frac{B^{+}_{\alpha} -
B^{-}_{\alpha}}{2} \Biggr) \frac{1}{\cos \theta_{\alpha}}.
\end{equation}

Physically, the constant area of the ellipse given by
\begin{equation}
A_{\alpha} = \frac{\pi}{\sqrt{\lambda^{+}_{\alpha}
\lambda^{-}_{\alpha}}} = \coth \Biggl(\frac{\beta
\omega_{\alpha,i}}{2} \Biggr),
\end{equation}
implies that the phase transition only squeezes the initial
quantum state of the massive scalar field unless there is a
mode-mixing (coupling). The entropy of each mode given by
\begin{equation} S_{\alpha} \approx \ln
A_{\alpha} = \ln \coth \Biggl(\frac{\beta \omega_{\alpha,i}}{2} \Biggr),
\end{equation}
is also constant, i.e. isentropic, and vanishes for the vacuum
state as expected. Without coarse graining the entropy of the
massive scalar field does not change.

\section{Classical Correlation of Long Wavelength Modes}

We now study classical correlation and quantum decoherence of the
density matrix (\ref{den mat}) or (\ref{den mat2}) and the Wigner
function (\ref{wigner}) or (\ref{wigner2}). The Wigner function
(\ref{wigner}) oscillates for an excited state and is not always
positive definite, so it does not describe a true probability
distribution in phase space. On the other hand, the Wigner
function is positive definite for the Gaussian vacuum and thermal
state. The thermal state is particularly of physical interest in
the phase transition because it is mostly likely that the system
starts from a thermal equilibrium.

A quantum system achieves quantum decoherence when the
interference between classical trajectories is lost, and it
recovers classical correlation when the Wigner function is peaked
along a classical trajectory. The coherence length $l_x$ of the
density matrix (\ref{den mat}) or (\ref{den mat2}) written in the
form
\begin{equation}
\rho_{T_{\alpha}} (\phi', \phi_{\alpha}) =
\Biggl(\frac{\Bigl\{\tanh \bigl(\beta \omega_{\alpha, i}/2
\bigr)\Bigr\}}{2\pi \varphi^*_{\alpha} \varphi_{\alpha}}
\Biggr)^{1/2} \exp \Bigl[- \Gamma_{\alpha,C} \phi_{\alpha, C}^2 -
\Gamma_{\alpha, \Delta} \phi_{\alpha, \Delta}^2 - \Gamma_{\alpha,
M} \phi_{\alpha, C} \phi_{\alpha, \Delta} \Bigr], \label{can den}
\end{equation}
is roughly determined by the width of $\phi_{\Delta, \alpha}^2$:
\begin{equation}
l_x = \frac{1}{\sqrt{\Gamma_{\alpha, \Delta}}} =
\left[\Biggl\{\tanh\Biggl(\frac{\beta \omega_{\alpha,i}}{2}\Biggr)
\Biggr\} \varphi_{\alpha}^* \varphi_{\alpha} \right]^{1/2}.
\end{equation}
Here the vacuum state result is $\tanh(\beta \omega_{\alpha, i}/2)
= 1$. Hence the coherence length increases or decreases depending
on whether $\varphi_{\alpha}$ grows or decays. A large coherence
length implies a large degree of quantum interference, so quantum
decoherence is conditioned by the small magnitude of $l_x$.

More rigorously, the representation-independent measure of
classical correlation of Ref. \cite{anderson} is given by
\begin{eqnarray}
\delta_{CC} &=& \sqrt{\frac{\Gamma_{\alpha, C}^2 \Gamma_{\alpha,
\Delta}^2}{\Gamma_{\alpha,M}^* \Gamma_{\alpha, M}}} \nonumber\\
&=& \frac{1}{4 (\varphi^*_{\alpha} \varphi_{\alpha} )^2 \Bigl|
\frac{d}{dt} \ln (\varphi_{\alpha}^* \varphi_{\alpha}) \Bigr|}.
\label{cc}
\end{eqnarray}
Similarly, the measure of quantum decoherence \cite{anderson},
which is determined by the ratio of the width of the off-diagonal
element to that of the diagonal element, is given by
\begin{eqnarray}
\delta_{QD} &=& \frac{1}{2} \sqrt{\frac{\Gamma_{\alpha,
C}}{\Gamma_{\alpha, \Delta}}} \nonumber\\ &=& \frac{1}{2} \Biggl\{
\tanh\Biggl(\frac{\beta \omega_{\alpha,i}}{2}\Biggr) \Biggr\}.
\label{qd}
\end{eqnarray}
This means that the massive scalar field preserves the same
initial quantum coherence throughout the phase transition.

The quantum system recovers classical correlation when
$\delta_{CC} \ll 1$  and loses quantum coherence when $\delta_{QD}
\ll 1$. So the system becomes classically correlated only when
$\varphi_{\alpha}$ grows during the evolution. In the opposite
case of decaying $\varphi_{\alpha}$, the degree of classical
correlation increases compared to that of quantum coherence, but
the system retains quantum coherence. To see how the classical
trajectory is recovered when $\delta_{CC} \ll 1$, we look into the
Wigner function (\ref{wigner2}). Each contour of the Wigner
function, which is quadratic $\pi_{\alpha}$ and $\phi_{\alpha}$,
depicts an ellipse in phase space. Suppose that both the classical
field $\phi_{\alpha}$ and auxiliary field $\varphi_{\alpha}$ grow
exponentially during the phase transition:
\begin{eqnarray}
 \left\{ \begin{array}{c} \phi_{\alpha} (t)
\\ \varphi_{\alpha} (t) \end{array} \right\}
\simeq c_{\alpha} e^{f_{\alpha} (t)},
\end{eqnarray}
then the classical trajectory obeys
\begin{equation}
\pi_{\alpha} (t) = \dot{\phi}_{\alpha} (t) \simeq
\dot{f}_{\alpha} (t) \phi_{\alpha} (t).
\end{equation}
On the other hand, the last term in Eq. (\ref{eff ham}) is
exponentially suppressed compared to the first two terms and the
overall coefficient of the exponent increases exponentially, so
the Wigner function (\ref{wigner2}) is  sharply peaked around a
trajectory in phase space:
\begin{equation}
\pi_{\alpha} = \Biggl[\frac{d}{dt} \ln(\varphi^*_{\alpha}
\varphi_{\alpha} )^{1/2} \Biggr] \phi_{\alpha} \simeq \dot{f}_{\alpha}
(t) \phi_{\alpha}.
\end{equation}
Indeed the Wigner function is sharply peaked around the classical
trajectory.

Now we apply the criterion on classicality to the second order
phase transition via an instantaneous quench, in which the mass
changes sign as
\begin{equation}
 m^2 (t) = \begin{cases}
  ~~m_i^2, &  t < 0, \\
  -m_f^2, & t > 0. \label{ins mass}
\end{cases}
\end{equation}
Before the quench, the most general solution to Eq. (\ref{cl sol})
is given by
\begin{equation}
\varphi_{\alpha, i} = \frac{1}{\sqrt{2\omega_{\alpha, i}}} \Biggl[
\cosh r_{\alpha} e^{- i \omega_{\alpha, i} t}
+ e^{- i \delta_{\alpha}} \sinh r_{\alpha} e^{i \omega_{\alpha, i} t}
\Biggr], \label{gen sol}
\end{equation}
where
\begin{equation}
\omega_{\alpha, i} = \sqrt{ m_i^2 + {\bf k}^2 }.
\end{equation}
The system is stable and has the minimum expectation value when
$r_{\alpha} = 0$, corresponding to the true vacuum state. As
$r_{\alpha} (\neq 0)$ represents a squeezing parameter of the true
vacuum state, the vacuum state in Eq. (\ref{num st}) is the
one-parameter squeezed vacua \cite{kim2}. After the quench, the
solution for each unstable long wavelength mode matches the
initial solution (\ref{gen sol}) continuously at the onset of the
quench $(t = 0)$ and is given by
\begin{eqnarray}
\varphi_{\alpha, U_f} &=& \frac{1}{\sqrt{2\omega_{\alpha, i}}} \Biggl[
(\cosh r_{\alpha} + e^{- i \delta_{\alpha}} \sinh r_{\alpha})
\cosh( \tilde{\omega}_{\alpha, f} t) \nonumber\\&&
- i( \cosh r_{\alpha} - e^{- i \delta_{\alpha}}
\sinh r_{\alpha}) \frac{\omega_{\alpha, i}}{\tilde{\omega}_{\alpha, f}}
\sinh( \tilde{\omega}_{\alpha, f} t) \Biggr], \label{uns sol}
\end{eqnarray}
where
\begin{equation}
\tilde{\omega}_{\alpha, f} = \sqrt{m_f^2 - {\bf k}^2}.
\end{equation}
Whereas the solution for each stable short wavelength mode is
given by
\begin{eqnarray}
\varphi_{\alpha, S_f} &=& \frac{1}{\sqrt{2\omega_{\alpha, i}}} \Biggl[
(\cosh r_{\alpha} + e^{- i \delta_{\alpha}} \sinh r_{\alpha})
\cos(\omega_{\alpha, f} t) \nonumber\\&&
- i( \cosh r_{\alpha} - e^{- i \delta_{\alpha}}
\sinh r_{\alpha}) \frac{\omega_{\alpha, i}}{\omega_{\alpha, f}}
\sin(\omega_{\alpha, f} t) \Biggr], \label{st sol}
\end{eqnarray}
where
\begin{equation}
\omega_{\alpha, f} = \sqrt{{\bf k}^2 - m_f^2}.
\end{equation}

Thus, after the quench, the amplitude square of the long
wavelength mode increases exponentially as
\begin{equation}
\varphi_{\alpha, U_f}^* \varphi_{\alpha, U_f} = \frac{1}{2
\tilde{\omega}_{\alpha, f}} [\cosh (2 r_{\alpha})
\cosh (2 \tilde{\omega}_{\alpha, f} t)  + \sin \delta_{\alpha}
\sinh (2 r_{\alpha}) \sinh (2 \tilde{\omega}_{\alpha, f}t)].
\end{equation}
Therefore, the measure of classical correlation (\ref{cc}) for the
long wavelength modes decreases exponentially to zero as the
spinodal instability continues. Also, the Wigner function
(\ref{wigner}) is sharply peaked around the classical trajectory $
\pi_{\alpha} \simeq \tilde{\omega}_{\alpha, f} \phi_{\alpha}$. On
the contrary, the short wavelength modes have oscillating,
bounded, amplitudes
\begin{equation}
\varphi_{\alpha, S_f}^* \varphi_{\alpha, S_f} = \frac{1}{2
\omega_{\alpha, f}} [\cosh (2 r_{\alpha})
\cos (2\omega_{\alpha, f} t)  + \sin \delta_{\alpha}
\sinh (2 r_{\alpha}) \sin (2\omega_{\alpha, f}t)].
\end{equation}
Thus the measure of classical correlation ({\ref{cc}) cannot
become small for the short wavelength modes. But the measure of
quantum decoherence (\ref{qd}) of the long and short wavelength
modes remains constant.

In summary, the unstable long wavelength modes of the massive
scalar field recover classical correlation during the phase
transition whereas the stable short wavelength modes do not.
However, both the long and short wavelength modes retain the same
quantum coherence throughout the phase transition.

\section{Quantum Decoherence due to Mode-Mixing}

The long wavelength modes of the massive scalar field model in
Sec. III  gain classical correlation during the second order phase
transition without achieving quantum decoherence. A more realistic
model of the quantum phase transition is provided by $\Phi^4$
theory with the Lagrangian density
\begin{equation}
{\cal L} =  \frac{1}{2} [\dot{\phi}^2 - (\nabla \phi)^2 ] -
\frac{1}{2} m^2 (t) \phi^2 - \frac{\lambda}{4!} \phi^4,
\label{self}
\end{equation}
where $m^2 (t)$ is assumed to change sign as in Eq. (\ref{ins
mass}). Upon decomposing the field into the Fourier mode (\ref{fo
mod}), the Hamiltonian obtained from Eq. (\ref{self}) takes the
form \cite{kim-khanna}
\begin{equation}
H (t) = \sum_{\alpha} \Biggl[\frac{1}{2} \pi_{\alpha}^2 (t) +
\frac{1}{2} {\omega}^2_{\alpha} (t)\phi_{\alpha}^2 (t) \Biggr] +
\frac{\lambda}{4!} \Biggl[ \sum_{\alpha} \phi_{\alpha}^4 (t)+ 3
\sum_{\alpha \neq \beta} \phi_{\alpha}^2 (t) \phi_{\beta}^2 (t)
\Biggr], \label{self ham}
\end{equation}
where $\omega^2_{\alpha}$ is given by Eq. (\ref{freq}). The last
coupled anharmonic terms prohibit any analytical solution. In the
Gaussian approximation of the Hartree-Fock \cite{boyanovsky} or
Liouville-von Neumann method \cite{kim-lee}, the number state of
each mode is still given by Eq. (\ref{osc wav}), where
$\varphi_{\alpha}$ now obeys the mean-field equation
\begin{equation}
\ddot{\varphi}_{\alpha} (t) + \Biggl[ \omega^2_{\alpha} (t) +
\frac{\lambda}{2} \sum_{\beta} \varphi^*_{\beta}(t)
\varphi_{\beta}(t) \Biggr] \varphi_{\alpha}(t) = 0. \label{self
eq}
\end{equation}
In the Gaussian approximation the term $\phi^4$ affects only the
auxiliary variable $\varphi_{\alpha}$ through Eq. (\ref{self eq}),
but does not change the form of the state (\ref{osc wav}).
Therefore, the Gaussian approximation for the self-interacting
scalar field (\ref{self}) does not lead to quantum decoherence. As
the short wavelength modes provide an environment to the long
wavelength modes, the mode-mixing is expected to lead to quantum
decoherence.

To analytically study the mode-mixing effect between a long and
short wavelength mode, we consider an exactly solvable model
\begin{equation}
H_{\rm M} (t) =  \frac{1}{2} \pi_{S}^2 + \frac{1}{2}
{\omega}^2_{S} (t) \phi_{S}^2 + \frac{1}{2} \pi_{H}^2 +
\frac{1}{2} {\omega}^2_{H} (t) \phi_{H}^2 + \lambda  \phi_{S}
\phi_{H}. \label{model ham}
\end{equation}
Here the subscript $S$ and $H$ denotes the long and short
wavelength mode, respectively. As $\omega^2_{H}$ and
$\omega^2_{H}$ depend on time through $m^2 (t)$, the orthogonal
transformation leading to normal modes depends time explicitly.
Hence the product of the instantaneous eigenstates of each normal
mode is not an exact quantum state of the original Hamiltonian
(\ref{model ham}). The Schr\"{o}dinger equation has a Gaussian
state of the form
\begin{equation}
\Psi_0 (\phi_{S}, \phi_{H}, t) = N(t) \exp \Bigl[- \{A_{S} (t)
\phi_{S}^2 + \lambda B (t) \phi_{S} \phi_{H} + A_{H} (t)
\phi_{H}^2 \} \Bigr],
\end{equation}
where $N$ is the normalization constant, which is not of much
concern to this paper. The coefficients determined by the
Schr\"{o}dinger equation are
\begin{eqnarray}
A_{S}  (t) &=& -i\frac{\dot{u}^* (t)}{2 u^* (t)}, \nonumber\\
A_{H} (t) &=& - i \frac{\dot{v}^* (t)}{2 v^* (t)}, \nonumber\\
B(t) &=& i \frac{\int u^* (t) v^* (t)}{u^*(t) v^*(t)},
\end{eqnarray}
where
\begin{eqnarray}
\ddot{u} (t) + \Biggl[\omega_{S}^2 (t) + \lambda^2
\Biggl\{\frac{\int u (t) v(t)}{u(t) v(t)} \Biggr\}^2 \Biggr] u (t)
= 0, \label{mx eq1}\\ \ddot{v} (t) + \Biggl[ \omega_{H}^2 (t) +
\lambda^2 \Biggl\{\frac{\int u (t) v(t)}{u(t) v(t)} \Biggr\}^2
\Biggr] v (t) = 0. \label{mx eq2}
\end{eqnarray}
We solve Eqs. (\ref{mx eq1}) and (\ref{mx eq2}) separately for two
cases: before and after the phase transition.

First, before the phase transition, the frequencies take the
constant values $\omega_{S}^2 (t) = \omega_{S, i}^2$ and
$\omega_{H}^2 (t) = \omega_{H, i}^2$. The solutions to Eqs.
(\ref{mx eq1}) and (\ref{mx eq2}) are found to be
\begin{eqnarray}
u (t) &=& \frac{1}{\sqrt{2 \Omega_S}} e^{- i \Omega_S t},
\nonumber\\ v(t) &=& \frac{1}{\sqrt{2 \Omega_H}} e^{- i \Omega_H
t},
\end{eqnarray}
where
\begin{eqnarray}
\Omega_S^2 &=& \omega_{S,i}^2 - \Biggl(\frac{\lambda}{\Omega_S +
\Omega_H } \Biggr)^2, \nonumber\\ \Omega_H^2 &=& \omega_{H,i}^2 -
\Biggl(\frac{\lambda}{\Omega_S + \Omega_H } \Biggr)^2.
\end{eqnarray}
Hence the Gaussian state is given by
\begin{equation}
\Psi_0 (\phi_S, \phi_H) = N \exp \Biggl[-
\Biggl\{\frac{\Omega_S}{2}\phi_S^2 + \frac{\lambda}{\Omega_S +
\Omega_H} \phi_S \phi_H +\frac{\Omega_H}{2}\phi_H^2
\Biggr\}\Biggr], \nonumber
\end{equation}
and the reduced density matrix for the long wavelength mode by
\begin{eqnarray}
\rho_{\rm R} (\phi_S', \phi_S) &=& \int_{- \infty}^{+ \infty} d
\phi_H \rho (\phi_S', \phi_H'; \phi_S, \phi_H) \nonumber\\ &=& N^*
N \sqrt{\frac{\pi}{\Omega_H}} \exp \Biggl[- \Biggl\{ \Omega_{S} -
\frac{\lambda^2}{\Omega_H (\Omega_S + \Omega_H)^2}  \Biggr\}
\phi_{S, C}^2 - \Omega_S \phi_{S, \Delta}^2 \Biggr].
\end{eqnarray}
Finally, from the coefficients of the density matrix (\ref{can
den})
\begin{equation}
\Gamma_{S,C} = \Omega_{S} - \frac{\lambda^2}{\Omega_H (\Omega_S +
\Omega_H)^2}, \quad \Gamma_{S,\Delta} = \Omega_S, \quad
\Gamma_{S,M} = 0,
\end{equation}
the measure of quantum decoherence is given by
\begin{eqnarray}
\delta_{QD} &=& \frac{1}{2}
\sqrt{\frac{\Gamma_{S,C}}{\Gamma_{S,\Delta}}} \nonumber\\ &=&
\frac{1}{2} \sqrt{1 - \frac{\lambda^2}{\Omega_S \Omega_H (\Omega_S
+ \Omega_H)^2}},\label{dec2}
\end{eqnarray}
and that of classical correlation by
\begin{equation}
\delta_{CC} = \sqrt{\frac{\Gamma_{S,C}^2
\Gamma_{S,\Delta}^2}{\Gamma_{S,M}^* \Gamma_{S,M}}}.\label{cor2}
\end{equation}
Therefore, quantum decoherence is achieved by the mode-mixing, but
the characteristic behavior of classical correlation does not
change.

Second, after the phase transition, Eqs. (\ref{mx eq1}) and
(\ref{mx eq2}) become
\begin{eqnarray}
\ddot{u} (t) + \Biggl[- \tilde{\omega}_{S, f}^2 + \lambda^2
\Biggl\{\frac{\int u (t) v(t)}{u(t) v(t)} \Biggr\}^2 \Biggr] u (t)
= 0 \nonumber\\ \ddot{v} (t) + \Biggl[\omega_{H, f}^2 + \lambda^2
\Biggl\{\frac{\int u (t) v(t)}{u(t) v(t)} \Biggr\}^2 \Biggr] v (t)
= 0.
\end{eqnarray}
In the weak coupling limit $M \ll \tilde{\omega}_{S, f},
\omega_{H, f}$, the approximate solutions are found to be
\begin{eqnarray}
u (t) = c_1 e^{\tilde{\Omega}_{S} t}, \quad v(t) = c_2 e^{- i
\Omega_H t},
\end{eqnarray}
where
\begin{eqnarray}
\tilde{\Omega}_S = \tilde{\omega}_{S, f} - \frac{\lambda^2}{2
\tilde{\omega}_{S, f}} \Biggl(\frac{\tilde{\omega}_{S, f} + i
\omega_{H, f}}{\tilde{\omega}_{S, f}^2 + \omega_{H, f}^2}
\Biggr)^2 - i \frac{\lambda^4}{2 \tilde{\omega}_{S, f}^2
\omega_{H, f}} \Biggl(\frac{\tilde{\omega}_{S, f} + i \omega_{H,
f}}{\tilde{\omega}_{S, f}^2 + \omega_{H, f}^2} \Biggr)^4 -
\frac{\lambda^4}{8 \tilde{\omega}_{S, f}^3}
\Biggl(\frac{\tilde{\omega}_{S, f} + i \omega_{H,
f}}{\tilde{\omega}_{S, f}^2 + \omega_{H, f}^2} \Biggr)^4,
\nonumber\\ \tilde{\Omega}_H = \omega_{H, f} + \frac{\lambda^2}{2
\omega_{H, f}} \Biggl(\frac{\tilde{\omega}_{S, f} + i \omega_{H,
f}}{\tilde{\omega}_{S, f}^2 + \omega_{H, f}^2} \Biggr)^2 + i
\frac{\lambda^4}{2 \tilde{\omega}_{S, f} \omega_{H, f}^2}
\Biggl(\frac{\tilde{\omega}_{S, f} + i \omega_{H,
f}}{\tilde{\omega}_{S, f}^2 + \omega_{H, f}^2} \Biggr)^4 -
\frac{\lambda^4}{8 \omega_{H, f}^3}
\Biggl(\frac{\tilde{\omega}_{S, f} + i \omega_{H,
f}}{\tilde{\omega}_{S, f}^2 + \omega_{H, f}^2} \Biggr)^4.
\end{eqnarray}
These solutions are accurate at late times provided that ${\rm Re}
(\tilde{\Omega}_S) t \gg 1$. Then, the reduced density matrix is
given by
\begin{equation}
\rho_{\rm R} (\phi_s', \phi_S) = N^* N \sqrt{ \frac{\pi}{A_H^* +
A_H}} \exp \bigl[- \Gamma_{S, C} \phi_{S, C}^2 - \Gamma_{S,
\Delta} \phi_{S, \Delta}^2 - \Gamma_{S,M} \phi_{S, C} \phi_{S,
\Delta} \bigr],
\end{equation}
where
\begin{eqnarray}
\Gamma_{S, C} &=& 2 \Biggl[ {\rm Re} A_S - \frac{\lambda^2 ({\rm
Re} B)^2}{4 {\rm Re} A_H} \Biggr], \\ \Gamma_{S, \Delta} &=& 2
\Biggl[ {\rm Re} A_S + \frac{\lambda^2 ({\rm Im} B)^2}{4 {\rm Re}
A_H} \Biggr],\\ \Gamma_{S,M} &=& 4 i \Biggl[ {\rm Im} A_S-
\frac{\lambda^2 ({\rm Re B}) (\rm Im B)}{ 2{\rm Re} A_H} \Biggr].
\end{eqnarray}
To order $\lambda^4$, the coefficients are approximated by
\begin{eqnarray}
\Gamma_{S, C} &=& \frac{5 \lambda^4 \omega_{H, f}
(\tilde{\omega}_{S, f} - \omega_{H, f})}{2 \tilde{\omega}_{S,f}
(\tilde{\omega}_{S, f}^2 + \omega_{H, f}^2)^4 }, \\ \Gamma_{S,
\Delta} &=& \frac{\lambda^2 }{\omega_{H,f} (\tilde{\omega}_{S,
f}^2 + \omega_{H, f}^2)} \Biggl[1 + \frac{\lambda^2 \omega_{H,
f}}{(\tilde{\omega}_{S, f}^2 + \omega_{H, f}^2)^3} \Biggl(
\frac{\tilde{\omega}_{S, f}^4}{2 \omega_{H, f}^3} -
\frac{3\tilde{\omega}_{S, f}^2}{\omega_{H, f}} - \frac{5
\omega_{H, f}^2 }{2 \tilde{\omega}_{S, f}} + \omega_{H, f} \Biggr)
\Biggr] ,
\\ \Gamma_{S,M} &=& - 2 i \tilde{\omega}_{S, f} \Biggl[1 + \frac{\lambda^2
\omega_{H, f}^2}{2 \tilde{\omega}_{S, f}^2 (\tilde{\omega}_{S,
f}^2 + \omega_{H, f}^2)^2} + \frac{\lambda^4}{(\tilde{\omega}_{S,
f}^2 + \omega_{H, f}^2)^4} \Biggl(\frac{16}{5} - \frac{3
\omega_{H, f}^2}{2 \tilde{\omega}_{S, f}^2} - \frac{\omega_{H,
f}^4}{4 \tilde{\omega}_{S, f}^4}\Biggr) \Biggr].
\end{eqnarray}
Therefore, to $\lambda^4$ order, the measure of quantum
decoherence is given by
\begin{equation}
\delta_{QD} = \frac{\lambda}{2} \sqrt{\frac{5\omega_{H,
f}^2(\tilde{\omega}_{S, f} - \omega_{H, f})}{2 \tilde{\omega}_{S,
f} (\tilde{\omega}_{S, f}^2 + \omega_{H, f}^2)^2}},
\end{equation}
and that of classical correlation by
\begin{equation}
\delta_{CC} = \frac{5\lambda^6}{4} \frac{\vert \tilde{\omega}_{S,
f} - \omega_{H, f} \vert}{\tilde{\omega}_{S, f}^2
(\tilde{\omega}_{S, f}^2 + \omega_{H, f}^2)^5}.
\end{equation}

A few comments are in order. First, quantum decoherence is
observed even for the quadratic Hamiltonian (\ref{model ham}). It
is a consequence of the mode-mixing and does not exclusively
pertain to the nonlinearity of the system. For the quadratic
system with constant parameters, the parameters introduced in Ref.
\cite{han} as
\begin{equation}
e^{\eta} = \sqrt{\frac{\lambda_+}{\lambda_-}}, \quad  \tan (2
\theta) = \frac{\lambda}{\omega_{H}^2 - \omega_{S}^2},
\end{equation}
where
\begin{equation}
\lambda_{\pm} = \frac{1}{2} \Biggl[\omega_S^2 + \omega_H^2 \pm
\sqrt{(\omega_S^2 - \omega_H^2)^2 + \lambda^2} \Biggr],
\end{equation}
yield another expression for quantum decoherence
\begin{equation}
\delta_{QD} = \frac{1}{2} \sqrt{\frac{1}{\cosh^2 \eta - \sinh^2
\eta  \cos^2(2 \theta)}}.
\end{equation}
Obviously, there is no quantum decoherence, $\delta_{QD} = 1/2$,
for the zero mode-mixing $\lambda = \theta = 0$. In the limiting
case of weak coupling $\lambda \ll \vert \omega_H^2 - \omega_S^2
\vert$, the mixing angle is small $(\theta \approx 0)$ and the
measure of quantum decoherence becomes $\delta_{QD} \approx 1/2$,
which is consistent with the result (\ref{dec2}). Whereas, for the
two identical oscillators $\omega_S = \omega_H$ with the maximum
mixing angle $\alpha = \pi/4$, the measure is simply given by
\begin{equation}
\delta_{QD} = \frac{1}{2 \cosh \eta}
\end{equation}
where
\begin{equation}
e^{\eta} = \sqrt{\frac{2 \omega_S^2 + \lambda}{2 \omega_S^2 -
\lambda}}.
\end{equation}
Second, it is shown that the unstable system efficiently achieves
quantum decoherence through the coupling with an environment. This
effect is also observed in the Bateman or the Feshbach-Tikochinsky
oscillator. The Bateman or Feshbach-Tikochinsky oscillator is a
conserved two-oscillator system which consists of a damped and an
amplified oscillator \cite{bateman,f-t}. The amplified oscillator
has an exponentially increasing amplitude at the expense of the
damped oscillator and reminds us of the unstable mode of the model
(\ref{model ham}). It is found that the amplified oscillator shows
a similar quantum decoherence \cite{ksk}. Thus it may be inferred
that even in quadratic systems quantum decoherence is a
consequence of the mode-mixing (coupling) of systems to the
environment.

\section{Quantal Order Parameter}

In the previous section it is found that the unstable long
wavelength modes recover their classical correlation but the
stable short wavelength modes retain the quantum coherence during
the quench. Thus the parametric interaction via the quench makes
the quantum phase transition exhibit classicality. As some, though
not all, of the essential features of the second order phase
transition can also be explained by classical theory, it would be
interesting to find a quantal quantity that behaves like a
classical order parameter.

In quantum theory the order parameter is frequently defined by
$\langle \hat{\phi} (t, {\bf x}) \rangle$, the mean value of the
field. The mean value of a coherent state satisfies the classical
field equation with quantum correction from fluctuations. However,
the thermal state or Gaussian vacuum studied in this paper is
symmetric, so the expectation value of the field modes (field)
with respect to this state vanishes:
\begin{equation}
\langle \hat{\phi}_{\alpha} \rangle_{\rm T} = {\rm Tr} (
\hat{\phi}_{\alpha} \hat{\rho}_{{\rm T}, \alpha}) = 0.
\end{equation}
The nonzero expectation values come from the quadratic moment of
the field or field modes. One candidate is the two-point function
$\langle \hat{\phi} (t, {\bf x}) \hat{\phi} (t, {\bf y}) \rangle$.
It satisfies the classical field equation in the case of a massive
scalar field but not in the case of a self-interacting scalar
field. As the two-point function is quadratic in the field
variable, we wish to define a quantity that is linear in the field
variable and satisfies the classical field equation.

We note that the quadratic moment of the field mode also has the
nonzero expectation value
\begin{equation}
\langle \hat{\phi}_{\alpha}^2 \rangle_{\rm T} = \varphi_{\alpha}^*
\varphi_{\alpha} \coth\Biggl(\frac{\beta \omega_{\alpha, i}}{2} \Biggr).
\end{equation}
The classical equation (\ref{cl sol}) for a complex $\varphi_{\alpha}$
leads to the equation \cite{kim-lee}
\begin{equation}
\ddot{\xi}_{\alpha} (t) + \omega_{\alpha}^2 (t) \xi_{\alpha} (t)
- \frac{1}{4 \xi_{\alpha}^3 (t)} = 0, \label{am eq}
\end{equation}
where $\xi_{\alpha}$ is the amplitude of complex
$\varphi_{\alpha}$:
\begin{equation}
\varphi_{\alpha} (t) = \xi_{\alpha} (t) e^{i \theta_{\alpha} (t)}.
\end{equation}
The last term in Eq. (\ref{am eq}), which originates from
quantization rule and guarantees the Heisenberg uncertainty
principle, can be neglected for the long wavelength modes because
$\xi_{\alpha}$ increases exponentially during the quench.
Therefore, it approximately satisfies
\begin{equation}
\ddot{\xi}_{\alpha} (t) + \omega_{\alpha}^2 (t) \xi_{\alpha} (t)
\simeq  0. \label{app eq}
\end{equation}
As each $\xi_{\bf k} e^{i {\bf k} \cdot {\bf x}}$ satisfies
the classical field equation, it follows that the quantal quantity defined by
\begin{eqnarray}
\phi_{\rm O} (t, {\bf x}) &=& \int_{k = 0}^{k \leq m_f} \frac{d^3
{\bf k}}{(2\pi)^3} \xi_{\bf k} (t) e^{i {\bf k} \cdot {\bf x}}
\nonumber\\ &=& \int_{k =0}^{k \leq m_f} \frac{d^3 {\bf
k}}{(2\pi)^3} \Biggl[\langle \hat{\phi}_{\alpha}^2 \rangle_{\rm T}
\tanh\Biggl(\frac{\beta \omega_{\alpha, i}}{2} \Biggr)
\Biggr]^{1/2} e^{i {\bf k} \cdot {\bf x}} \label{or pa}
\end{eqnarray}
also satisfies the classical field equation
\begin{equation}
\ddot{\phi}_{\rm O} (t, {\bf x})  - \nabla^2 \phi_{\rm O} (t, {\bf x})
+ m^2 (t) \phi_{\rm O} (t, {\bf x})  \simeq 0.
\end{equation}
It should be remarked that $\phi_{\rm O}$ is linear in the field
variable and is defined out of the long wavelength modes, which
become classically correlated during the quench. In this sense
$\phi_{\rm O}$ carries many features of the classical order
parameter.

\section{Conclusion}

We studied classical correlation and quantum decoherence in
quantum phase transition. For the initial state such as the
Gaussian vacuum or number state or thermal equilibrium, the
expectation value of the field vanishes throughout the phase
transition. Hence it cannot be used as the order parameter. The
phase transition in such a symmetric state, however, can be
explained entirely within the framework of quantum field theory.
In the massive scalar field model the long wavelength modes become
unstable and grow exponentially during the quench. The
quantitative measure from the density matrix shows classical
correlation of the long wavelength modes during the phase
transition. Likewise, the Wigner functions for these modes are
sharply peaked around their classical trajectories and thus
confirm classical correlation. Whereas, the short wavelength modes
are stable throughout the quench process and retain the quantum
coherence.

Classicality is, in a strict sense, conditioned not only by the
recovery of classical correlation but also by the loss of quantum
coherence. However the massive scalar model does not achieve
completely quantum decoherence, because the measure of quantum
decoherence keeps the value determined by the initial Gaussian
vacuum or thermal equilibrium. Hence the system recovers
classicality partially, depending on the relative magnitude
between the measures of classical correlation and quantum
decoherence. To achieve quantum decoherence it is necessary for
the long wavelength modes to couple to an environment. The
environment-induced decoherence was studied numerically in the
self-interacting scalar field model, where the long wavelength
modes are coupled to the short wavelength modes
\cite{lombardo,lombardo2}. However, it is very hard to find any
analytical solution of the self-interacting scalar model. The
Gaussian approximation of the Hartree-Fock \cite{boyanovsky} or
the Liouville-von Neumann method \cite{kim-lee} does not include
the mode-mixing effect appropriately to explain quantum
decoherence.

In this paper, to analytically study the mode-mixing effect, we
turned on a linear coupling between the long and short wavelength
modes of the massive scalar field. This exactly solvable model
shows that the long wavelength modes indeed decohere because of
the mode-mixing with the short wavelength modes. Therefore,
classicality emerges during the phase transition from the unstable
long wavelength modes coupled to the stable short wavelength
modes. These long wavelength modes behave classically and provide
a quantal quantity that is linear in the field variable, satisfies
the classical field equation, and thus behaves like a classical
order parameter.

After this paper was completed, we were informed of Ref.
\cite{lombardo3}, which also shows quantum decoherence of the long
wavelength modes through the nonlinear coupling to the short
wavelength modes in the self-interacting scalar field and confirms
the numerical result \cite{lombardo}. However, any analytical
solution in a closed form has not been found for the
self-interacting scalar field. At best, the analytical solution
can be found by perturbatively including the mode-mixing terms
\cite{kim-khanna}, which is beyond the scope of this paper. The
non-Gaussian effect from the mode-mixing is expected to result in
quantum decoherence. As quantum decoherence is largely a
consequence of mode-mixing (coupling), our exactly solvable model
may shed some light in understanding the mechanism of
classicality.

\begin{acknowledgments}

S.P.K. would like to thank F.C. Khanna and S. Sengupta for useful
discussions on the mode-mixing between the long and short
wavelength modes and Prof. Y.S. Kim for useful information. Also
he was deeply indebted to D.N. Page for helpful discussions on
quantum decoherence and comments. The authors would like to
express their appreciation for the warm hospitality of the
Theoretical Physics Institute, University of Alberta. This work
was supported in part by KOSEF under Grant No. 1999-2-112-003-5
and by NSERC of Canada.

\end{acknowledgments}


\end{document}